%% file: main.tex
\documentclass[preprint,12pt]{elsarticle}

\usepackage{amssymb}
\usepackage{amsmath}
\usepackage{url}
\usepackage{listings}
\usepackage{xcolor}
\usepackage[colorinlistoftodos]{todonotes}
\usepackage{enumitem}

\lstdefinelanguage{JavaScript}{
  keywords={typeof, new, true, false, catch, function, return, null, catch, switch, var, if, in, while, do, else, case, break, const},
  keywordstyle=\color{blue}\bfseries,
  ndkeywords={class, export, boolean, throw, implements, import, this, it, expect, test, require},
  ndkeywordstyle=\color{darkgray}\bfseries,
  identifierstyle=\color{black},
  sensitive=false,
  comment=[l]{//},
  morecomment=[s]{/*}{*/},
  commentstyle=\color{purple}\ttfamily,
  stringstyle=\color{red}\ttfamily,
  morestring=[b]',
  morestring=[b]"
}

\journal{Science of Computer Programming}

\begin{document}

\begin{frontmatter}

\title{JS-TOD: Detecting Order-Dependent Flaky Tests in Jest} 

\author[1]{Negar Hashemi}
\author[1]{Amjed Tahir}
\author[2]{Shawn Rasheed}
\author[3]{August Shi}
\author[1]{Rachel Blagojevic}

\affiliation[1]{organization={Massey University}, country={New Zealand}}
\affiliation[2]{organization={ Universal College of Learning}, country={New Zealand}}
\affiliation[3]{organization={The University of Texas at Austin},
            country={United States}}

\begin{abstract}
We present \texttt{JS-TOD} (\underline{J}ava\underline{S}cript \underline{T}est \underline{O}rder-dependency \underline{D}etector), a tool that can extract, reorder, and rerun Jest tests to reveal possible order-dependent test flakiness.
Test order dependency is one of the leading causes of test flakiness. Ideally, each test should operate in isolation and yield consistent results no matter the sequence in which tests are run. However, in practice, test outcomes can vary depending on their execution order.
JS-TOD employed a systematic approach to randomising tests, test suites, and \texttt{describe} blocks. The tool is highly customisable, as one can set the number of orders and reruns required (the default setting is 10 reorder and 10 reruns for each test and test suite). Our evaluation using JS-TOD reveals two main causes of test order dependency flakiness: shared files and shared mocking state between tests.
\end{abstract}


\end{frontmatter}

\input{introduction}

\input{approach}
\input{using}

\input{evaluation}


\bibliographystyle{IEEEtran}
\bibliography{bibliography.bib}

\end{document}

%% file: introduction.tex
\section{Motivation }
Test flakiness is a significant issue in software testing. Flaky tests are known to impact product correctness and quality negatively \cite{luo2014empirical,amjed2022review,costa2022test}. Among the many causes of test flakiness, test order dependency is widely acknowledged as a common cause of test flakiness across multiple languages and application domains \cite{lam2019deflaker, gruber2021empirical}. 


Listing~\ref{lst:exp} shows an example of order-dependent tests in Jest. The tests check how many times the \texttt{logger.log} function was called. The first test (\texttt{`calls logger once'}) makes a call and expects it to be logged once, while the second test (\texttt{`logger has not been called yet'}) assumes no calls have been made yet. Since the mock keeps its state between tests, running the first test before the second causes the second one to fail. To keep things isolated, the mock should be cleared before each test.

\begin{lstlisting}[caption={Order-dependent tests in Jest},label={lst:exp}]
test('calls logger once', () => {
  logger.log(`Test Log`);
  expect(logger.log).toHaveBeenCalledTimes(1);
});

test('logger has not been called yet', () => {
  expect(logger.log).not.toHaveBeenCalled(); 
});
 \end{lstlisting}

There is some tooling support to automatically detect possible test order-dependent tests for Java (JUnit)~\cite{lam2019idflakies,li2022evolution} and Python (pytest)~\cite{gruber2023flapy,wang2022ipflakies}.  To the best of our knowledge, there are no similar tools for JavaScript (Jest).  

\
Below we present our Jest test-order dependency detection approach, implemented in our tool \texttt{JS-TOD}.

%% file: approach.tex
\section{Approach}

\subsection{Jest Overview}
Jest\footnote{\url{https://jestjs.io/}} is one of the most used testing frameworks in JavaScript~\cite{yost2023finding,taleb2023frameworks}. In Jest, test files (test suites) are structured using blocks with the following keywords: \texttt{describe}, \texttt{test} (or \texttt{it}). Test files also utilize test hooks, such as \texttt{beforeEach} and \texttt{afterAll}, which define when test fixtures and cleaning of test state should occur. These blocks help organise tests and manage setup/teardown logic. The \texttt{describe} block is used to group related tests together in the same block. By default, Jest runs tests in parallel. It runs test blocks in the order they appear in the file (first one first). However, it also offers multiple options to change the running order of tests or only run a subset of tests. For example, the \texttt{randomize}\footnote{\url{https://jestjs.io/docs/cli#--randomize}} option randomises the running order of tests within a test file.

\subsection{\texttt{JS-TOD} Implementation}
Similar to tools such as iDFlakies~\cite{lam2019idflakies}, iFixFlakies~\cite{Shi2019iFixFlakies}, and FlaPy~\cite{gruber2023flapy}, \texttt{JS-TOD} reveals order-dependent tests by executing test suites in different orders and recording the outcomes. While prior tools available on other languages often focus on classifying or automatically repairing order-dependent tests, \texttt{JS-TOD} emphasizes a lightweight detection approach of test order dependency in Jest.

\texttt{JS-TOD} reveals order-dependent behaviour by extracting, reordering, and rerunning tests, based on a user-defined number of permutations and reruns. It does not perform automated classification or repair, leaving the investigation of root causes to developers. Jest's default \texttt{randomize} option changes the execution order of tests within a test suite, including individual tests and \texttt{describe} blocks based on a seed value. Using the same seed ensures the order is reproducible across runs. In contrast, \texttt{JS-TOD} performs systematic test reordering at three different levels: test suites (files), \texttt{describe} blocks, and test blocks. Users can specify the level of reordering, the number of permutations to generate, and the number of reruns. \texttt{JS-TOD} also saves the newly generated test orders as separate test suites for deeper analysis. The tool is publicly available on GitHub\footnote{\url{https://github.com/Negar-Hashemi/JS-TOD}}.

\begin{figure}[!htp]
    \centering
    \includegraphics[width=\linewidth]{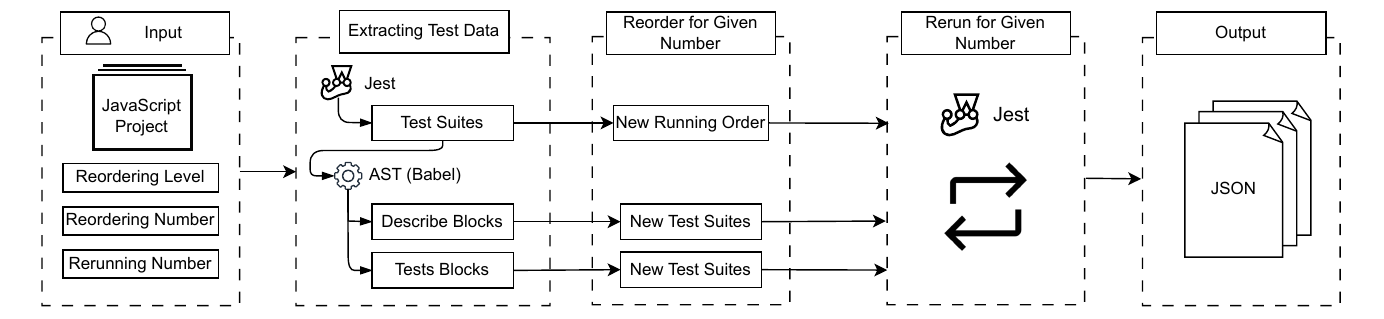}
    \caption{JS-TOD Approach}
    \label{fig:approach}
\end{figure}

Figure~\ref{fig:approach} illustrates \texttt{JS-TOD} approach. For a given project with Jest test files,
\texttt{JS-TOD} first extracts the test data of a project based on the specified reordering level. It then reorders and reruns the newly added test files for the given numbers. The result of each rerun is saved in a JSON file.

\subsubsection{Extracting Test data}
For automatically extracting the test suites' paths, we used Jest's  \texttt{-listTests} option\footnote{\url{https://jestjs.io/docs/cli\#--listtests}}, which allows us to list all test files that Jest will run. To enable reordering of \texttt{describe} blocks or individual tests, \texttt{JS-TOD} utilises the Babel toolchain\footnote{\url{https://babel.dev}}, a JavaScript compiler and source code transformer, to parse each test suite. Babel constructs an Abstract Syntax Tree (AST) for each suite, capturing its structure, parameters, and node types. Babel defines a range of node types, and by analysing the AST, we identify all \texttt{describe} blocks and tests present in each test suite of a given project. These blocks are detected by locating AST nodes where the type is ``identifier'' and the name is ``describe'' for \texttt{describe} blocks or either ``it'' or ``test'' for \texttt{test}s.

\subsubsection{Reordering Tests}
Using the extracted test data for a given project, \texttt{JS-TOD} can reorder tests based on the specified level. At the test level, \texttt{JS-TOD} reorders the tests within each \texttt{describe} block a given number of times and creates a new version of the test file in the same directory. The new test file retains the original filename with the reorder number appended to it.

Similarly, at the describe level, \texttt{JS-TOD} reorders the \texttt{describe} blocks within a file for the specified number of reorders. For each reorder, it generates a new test file named after the original, with the term describe and the reorder number appended.

At the test suite level, \texttt{JS-TOD} reorders the execution order of test files a given number of times and passes each permutation to \texttt{customSequencer.js}, a custom sequencer that extends Jest’s built-in testSequencer.

\subsubsection{Rerunning Tests}
After reordering the tests, \texttt{JS-TOD} reruns the tests for the specified number of times. It then saves the results in a directory under the project folder, depending on the reordering level: \texttt{\_extracted results\_} for tests, \texttt{\_extracted results describes\_} for \texttt{describe} blocks, and \texttt{\_extracted results test files\_} for test suites.

For test and \texttt{describe} block level reruns, the result files are named starting with \textit{testOutput}, followed by the name of the test suite and the rerun count (e.g., for a test suite named Foo with one rerun, the output file will be \texttt{testOutputFoo1}). For test suite-level reruns, the result files are named starting with \textit{testOutput}, followed by the reorder number and rerun count.

%% file: using.tex
\section{Using JS-TOD in CI/CD Pipelines}

To integrate \texttt{JS-TOD} into a CI/CD workflow, first navigate to the directory containing the tool:

\begin{lstlisting}[language=bash,numbers=none]
    cd/path/to/directory_containing_JS-TOD
\end{lstlisting}

Next, choose the desired level of reordering—tests, \texttt{describe} blocks, or entire test suites. To reorder and execute the tests of a specified project, use the following command:

\begin{lstlisting}[language=bash,numbers=none]
    node reorderRunner.js --project_path="/path/to/project" --rerun=<value> --reorder=<value>
\end{lstlisting}

\begin{itemize}
  \item \texttt{project\_path} sets the root directory of the target project.
  \item \texttt{rerun} specifies how many times each reordered configuration should be executed. The default is 10.
  \item \texttt{reorder} defines how many different reorders will be generated and tested. The default is also 10.
\end{itemize}

%% file: evaluation.tex
\section{Evaluation}

We used \texttt{JS-TOD} for reordering and rerunning tests in the study on order-dependent tests in JavaScript~\cite{hashemi2025detecting}. In this experiment, \texttt{JS-TOD}'s accuracy rate for returning the correct test paths was 90\% (i.e., 73 out of 81 programs returned the correct test paths) 
and 85\% for correctly reordering tests (i.e., 57 out of 67 programs were correctly reordered). By correctly reordering, we mean that \texttt{JS-TOD} recognises test blocks (including nested and individual tests) and reorders them without omitting or modifying any other parts of the test file. All reordering and rerunning steps were executed entirely within \texttt{JS-TOD}’s automated workflow.

\section{Limitations}
\texttt{JS-TOD} extracts test file paths using Jest's \texttt{--listTests} option, which is available starting from Jest version 20.0.0\footnote{\url{ https://github.com/jestjs/jest/releases/tag/v20.0.0}}. This option allows \texttt{JS-TOD} to systematically discover and process all test files in a project, regardless of their structure. However, this introduces a limitation: projects using Jest versions older than 20.0.0 do not support the \texttt{--listTests} option and, therefore, cannot use \texttt{JS-TOD} in its current form. 

Additionally, \texttt{JS-TOD} depends on modern ECMAScript features and Jest’s programmatic APIs (e.g., \texttt{testSequencer}), which may differ across major releases. Although this is unlikely in controlled environments, differences in Node.js or Jest versions may lead to reduced functionality or occasional execution failures. This is especially true in CI pipelines that use different base images or cached dependencies. These issues can be easily resolved by using an environment similar to the one validated in our evaluation (Node.js 18.16.1, npm 9.5.1, and Jest version 27 or higher).

Another limitation of \texttt{JS-TOD} is that reruns are executed sequentially. If a developer configures \texttt{JS-TOD} to perform many reorderings or reruns, the process may become time-consuming, particularly for large-scale projects with extensive test suites. 
\newline 

\noindent \textbf{\texttt{JS-TOD} Tool}: \url{https://github.com/Negar-Hashemi/JS-TOD}\\

\noindent \textbf{Dataset}: \url{https://doi.org/10.5281/zenodo.13852085}


